\UseRawInputEncoding
\documentclass[prl,aps,twocolumn]{revtex4}

\usepackage{graphicx}
\usepackage{amsmath,empheq}
\usepackage{amssymb}
\usepackage{ifthen}
\usepackage{booktabs}

\usepackage{graphicx}
\usepackage{dcolumn}
\usepackage{bm}
\usepackage{longtable}
\usepackage{gensymb}

\newcommand{\bb}{\begin{equation}}
\newcommand{\ee}{\end{equation}}
\newcommand{\ba}{\begin{eqnarray*}}
\newcommand{\ea}{\end{eqnarray*}}
\newcommand{\rhor}{\rho({\bf r})}
\newcommand{\dd}{{\rm d}}
\newcommand{\rr}{{\mathbf r}}

\begin{document}

\title{Breaking Cassie's law for condensation in a nano-patterned slit}

\author{Martin \surname{L\'aska}}
\affiliation{
{Department of Physical Chemistry, University of Chemical Technology Prague, Praha 6, 166 28, Czech Republic;}\\
 {The Czech Academy of Sciences, Institute of Chemical Process Fundamentals,  Department of Molecular Modelling, 165 02 Prague, Czech Republic}}                
\author{Andrew O. \surname{Parry}}
\affiliation{Department of Mathematics, Imperial College London, London SW7 2BZ, UK}
\author{Alexandr \surname{Malijevsk\'y}}
\affiliation{ {Department of Physical Chemistry, University of Chemical Technology Prague, Praha 6, 166 28, Czech Republic;}
 {The Czech Academy of Sciences, Institute of Chemical Process Fundamentals,  Department of Molecular Modelling, 165 02 Prague, Czech Republic}}

\begin{abstract}
\noindent We study the phase transitions of a fluid confined in a capillary slit made from two adjacent walls each of which are a periodic composite
of stripes of two different materials. For wide slits the capillary condensation occurs at a pressure which is described accurately by a combination
of the Kelvin equation and the Cassie law for an averaged contact angle. However, for narrow slits the condensation occurs in two steps involving an
intermediate bridging phase, with the corresponding pressures described by two new Kelvin equations. These are characterised by different contact
angles due to interfacial pinning, with one larger and one smaller than the Cassie angle. We determine the triple point and predict two types of
dispersion force induced Derjaguin-like corrections due to mesoscopic volume reduction and the  singular  free-energy contribution from nano-droplets
and bubbles. We test these predictions using a fully microscopic density functional model which confirms their validity even for molecularly narrow
slits. Analogous mesoscopic corrections are also predicted for two dimensional systems arising from thermally induced interfacial wandering.
\end{abstract}

\maketitle

\begin{figure}
\includegraphics[width=8cm]{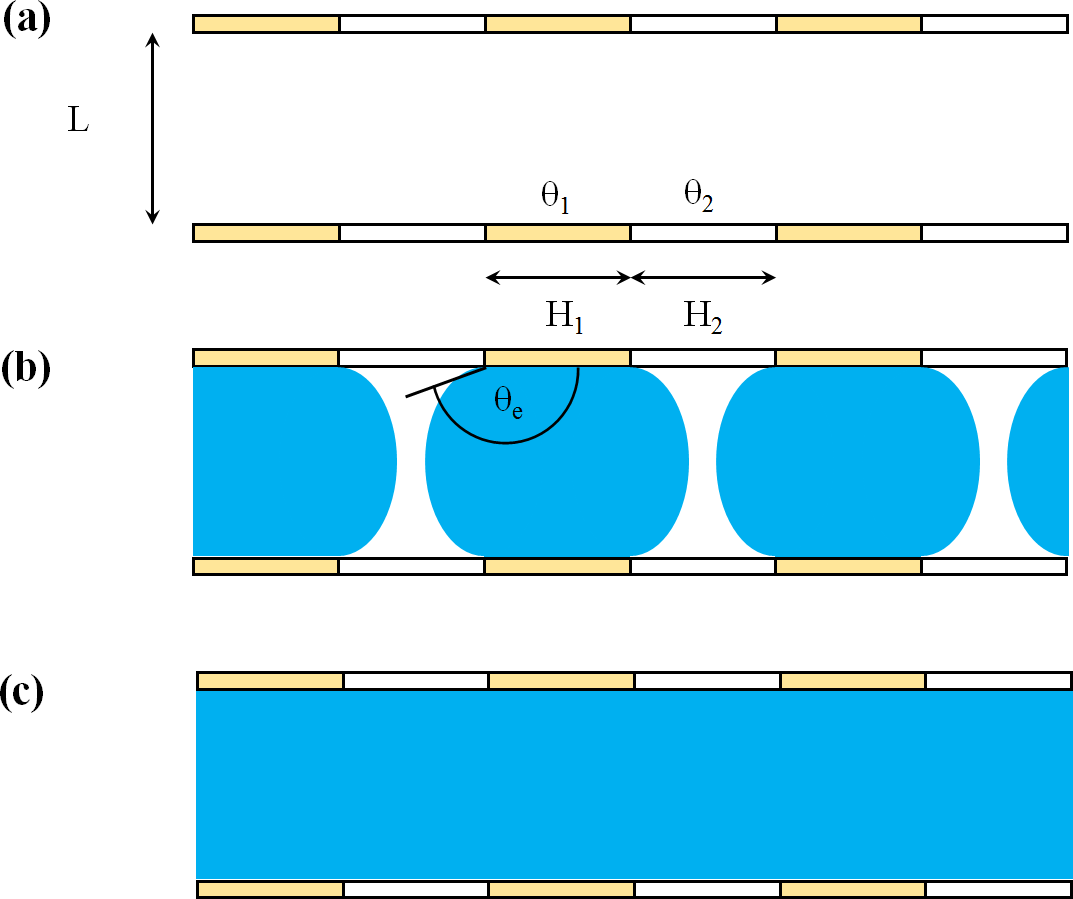}
\caption{Schematic illustration of a) gas-like, b) liquid bridge (blue)  and c) liquid-like  configurations in a periodic slit. The edge contact
angle $\theta_e$ is highlighted, drawn here for $p>p_{\rm sat}$.}
\end{figure}

The equilibrium phases of confined fluids have been the subject of long-standing interest. At the bulk critical point, fluctuations lead to the
thermal analogue of the Casimir force \cite{hertlein}, while at lower temperatures  the liquid-gas phase boundary is shifted leading to the
phenomenon of capillary condensation \cite{fisher, evans1, evans2, hend}. By having walls which preferentially adsorb different phases one can also
radically alter the nature of the phase equilibria due to interfacial effects \cite{parry_cwalls, binder03}. Indeed, it is now possible to observe
experimentally  the condensation in pores of different geometries fabricated using electron beam nanolitography \cite{bruschi}. In the present paper
we discuss the nature of capillary condensation in a slit for which the walls are periodically patterned with two different types of material. This
is much richer than that occurring for chemically homogeneous slits since the condensation from gas to liquid may either occur directly or in
two-steps via an intermediate bridge-like phase. The location of these phase transitions are described by generalized Kelvin equations which involve
one of three possible contact angles each associated with meniscus pinning. Only one of these angles is the Cassie angle \cite{cassie}, which lies
between the other two, somewhat analogous to contact angle hysteresis \cite{degennes, quere}. By allowing for dispersion forces we predict
Derjaguin-like corrections to the Kelvin equations due to volume reduction and the singular free-energy of droplets and bubbles adsorbed at the
walls.

To begin, we recall the macroscopic Kelvin equation and mesoscopic Derjaguin correction for condensation in a chemically homogeneous slit, made from
two infinite planar walls separated by a distance $L$. The fluid is at pressure $p$ (or equivalently chemical potential $\mu$), at a temperature $T$
below the critical point. Confinement changes the liquid-gas phase boundary, that is the pressure when gas condenses to liquid, away from the bulk
saturation curve $p_{\rm sat}(T)$. Macroscopically, the shift from $p_{\rm sat}$, at which capillary condensation (cc) occurs is described by the
Kelvin equation \cite{thomson}
\begin{equation}
\delta p_{cc}=\frac{2\gamma\cos\theta}{L}\,, \label{kelvin}
\end{equation}
where $\theta$ is the contact angle defined by Young's equation $\gamma_{wg}=\gamma_{wl}+\gamma\cos\theta$ for each semi-infinite wall. Here
$\gamma_{wg}$, $\gamma_{wl}$ and $\gamma$ are the tensions of the wall-gas, wall-liquid and liquid-gas interfaces respectively. For partial wetting
the Kelvin equation is remarkably accurate for $L$ down to tens of molecular length scales. However, for complete wetting ($\theta=0$), corrections
are apparent at the mesoscopic scale. In particular, for systems with dispersion forces the Kelvin equation is modified  to \cite{derj, evans85}
\begin{equation}
\delta p_{cc}=\frac{2\gamma}{L-3\ell}\,, \label{Derjaguin}
\end{equation}
which includes the Derjaguin correction allowing for the thickness $\ell$  of the liquid layer adsorbed at each wall in the gas-like phase. This is
very well approximated as $\ell=(2A/\delta p)^{-1/3}$ , where $A$ is the Hamaker constant and $\delta p=p_{\rm sat}-p$, corresponding to the wetting
layer thickness at a single wall \cite{lipowsky85, dietrich}.

We now turn  to a heterogeneous slit where the walls are made of two materials arranged into a periodic array of stripes of width $H_1$ and $H_2$
each characterised by different contact angles $\theta_1$ and $\theta_2$. The two walls are adjacent with translation invariance assumed along the
stripes (i.e. the stripes on the opposing walls are exactly aligned). We assume that material 1 preferentially adsorbs liquid relative to material 2
so that $\theta_1<\theta_2$.  The condensation in this capillary may happen by two mechanisms similar to that if there is a single stripe on each
surface \cite{laska} or between geometrically structured walls \cite{chmiel, rocken}. For wide slits it occurs via a single first-order phase
transition from a gas-like to liquid-like phase similar to that for a chemically homogeneous slit. However, if the slit is sufficiently narrow,
condensation happens in two steps via an intermediate phase in which liquid bridges locally condense between the stripes of the more wettable
material 1 (see Fig.~1). We first derive the generalised, macroscopic Kelvin equations which determine the phase boundaries for each type of
condensation:

{\bf{One-step condensation}}. The pressure at which gas condenses to liquid is determined by balancing the grand potential $\Omega$ per unit length
(along the stripe) and over a single period of the two phases. For the gas-like phase $\Omega_g\approx
-pHL+2\gamma_{wg}^{(1)}H_1+2\gamma_{wg}^{(2)}H_2$, where $\gamma_{wg}^{(i)}$ are the wall-gas tensions for each material and $H=H_1+H_2$. Similarly,
for the liquid-like phase we have $\Omega_l\approx  -p^\dagger HL+2\gamma_{wl}^{(1)}H_1+2\gamma_{wl}^{(2)}H_2$ where $p^\dagger$ is the pressure of
the metastable bulk liquid and $\gamma_{wl}^{(i)}$ are two wall-liquid surface tensions. Balancing the grand potentials determines that the value of
$p-p^\dagger\approx \delta p_{cc}$ at which (single step) capillary condensation occurs is given by
\begin{equation}
\delta p_{cc}=\frac{2\gamma\cos\theta_{\rm cas}}{L}\,, \label{kelvin_cassie}
\end{equation}
which is unaltered if the stripes are not perfectly adjacent, i.e. if the upper wall (say) in Fig.~1. is shifted or indeed, rotated so the stripes
are no longer parallel. This is the obvious generalization of the standard Kelvin equation and identifies that the appropriate contact angle
appearing in it is the familiar Cassie angle \cite{cassie},
\begin{equation}
\cos\theta_{\rm cas}=f_1\cos\theta_1+ f_2\cos\theta_2\,, \label{cassie}
\end{equation}
where $f_i=H_i/H$ is fraction of the wall area occupied by each material. Viewed in terms of phase separation, $\theta_{\rm cas}$ is the angle that a
circular meniscus of radius $R=\gamma/\delta p_{cc}$, which separates the coexisting phases, meets both walls at one of the edges between the two
materials.

{\bf{Two-step condensation}}. In this case, there are two phase boundaries corresponding to a pressure shift $\delta p_{gb}$, where the gas phase
locally condenses to a bridge phase, and a second $\delta p_{lb}$ (at higher pressure) when the bridge phase condenses to liquid. These are
determined by matching the grand potential of a bridge phase $\Omega_b$ with  $\Omega_g$ and $\Omega_l$. In each unit cell the liquid bridge is
bounded by two circular menisci of radius $R=\gamma/\delta p$ which are pinned at the edges between the two materials (see Fig.~1). These meet the
walls at an edge contact angle $\theta_e$ which is related to the pressure by $\cos\theta_e=L/2R$. The grand potential of the bridge phase is given
by $\Omega_b\approx -pA_g-p^\dagger A_l+2(H_1\gamma_{wl}^{(1)}+H_2\gamma_{wg}^{(2)})+2\gamma l_{\rm men}$ where $A_g$ and $A_l$ are the areas
occupied by the gas and (metastable) liquid, respectively, and $l_{\rm men}$ is the length of each meniscus. Equating this value with $\Omega_g$ and
$\Omega_l$ determines the pressure shifts  as
\begin{equation}
\delta p_{gb}=\frac{2\gamma\cos\theta_{gb}}{L}\,,\; \delta p_{lb}=\frac {2\gamma\cos\theta_{lb}}{L}\,,
\end{equation}
where $\theta_{gb}$ are $\theta_{lb}$ are the values of the edge contact angles at the respective phase transitions given by \cite{fin_slit}
 \bb
\cos\theta_1=\cos\theta_{gb}+\frac{L}{2H_1}\left [\sin\theta_{gb}+\sec\theta_{gb}\left(\frac{\pi}{2}-\theta_{gb}\right)\right] \label{bg_theta}
 \ee
 and
  \bb
\cos\theta_2=\cos\theta_{lb}-\frac{L}{2H_2}\left [\sin\theta_{lb}+\sec\theta_{lb}\left(\frac{\pi}{2}-\theta_{lb}\right)\right]\,, \label{bl_theta}
 \ee
which satisfy $\theta_{gb}<\theta_{\rm cas}<\theta_{lb}$. The phase transition occurring at $\delta p_{lb}$ is equivalent to the local evaporation of
liquid between the less wettable stripes as $p$ is reduced.

\begin{figure}
\includegraphics[width=8cm]{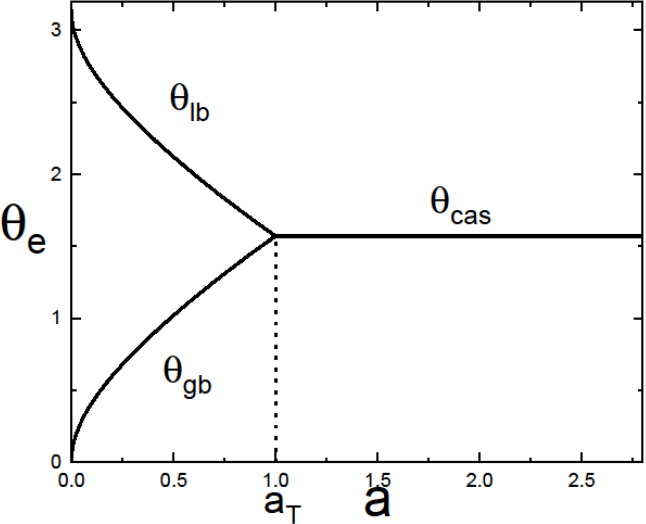}
\caption{Contact angles appearing in the macroscopic Kelvin equations for one step and two step condensation [see equations (3) and (5)]  in a
maximum contrast slit ($\theta_1=0$ and $\theta_2=\pi$)  as a function of the aspect ratio $a=L/H_1$. For equal area fractions,
$f_1=f_2=\frac{1}{2}$, the triple point occurs at $a_T=1$.}
\end{figure}

Condensation occurs via one step if $\theta_{gb}>\theta_{\rm cas}$ (wide slits) or two steps if $\theta_{gb}<\theta_{\rm cas}$ (narrow slits). The
marginal case between these two different mechanisms occurs when $\theta_{gb}=\theta_{lb}$ (equivalent to each being equal to $\theta_{\rm cas}$) and
identifies the triple point (T) where the gas-like, liquid-like and bridging phases coexist. This happens when the aspect ratio $a\equiv L/H_1$ takes
the value
\begin{equation}
a_T=\frac{2f_2(\cos\theta_1-\cos\theta_2)}{\sin\theta_{\rm cas}+\sec\theta_{\rm cas}\left(\frac{\pi}{2}-\theta_{\rm cas}\right)}\,. \label{aT}
\end{equation}
In the limit $f_1=0$, Eq.~(8) reduces to the result pertinent to a parallel plate geometry in which there is just a single, adjacent stripe, of
material 1 on each wall \cite{laska}. If the aspect ratio is greater than this value the condensation occurs via one step. We illustrate this in
Fig.~2 where we plot the three contact angles appearing in the generalised Kelvin equations for a \emph{maximum contrast} slit ($\theta_1=0$ and
$\theta_2=\pi$) as a function of the aspect ratio $a$ for equal stripe widths $H_1=H_2$.

\begin{figure}
\includegraphics[width=8cm]{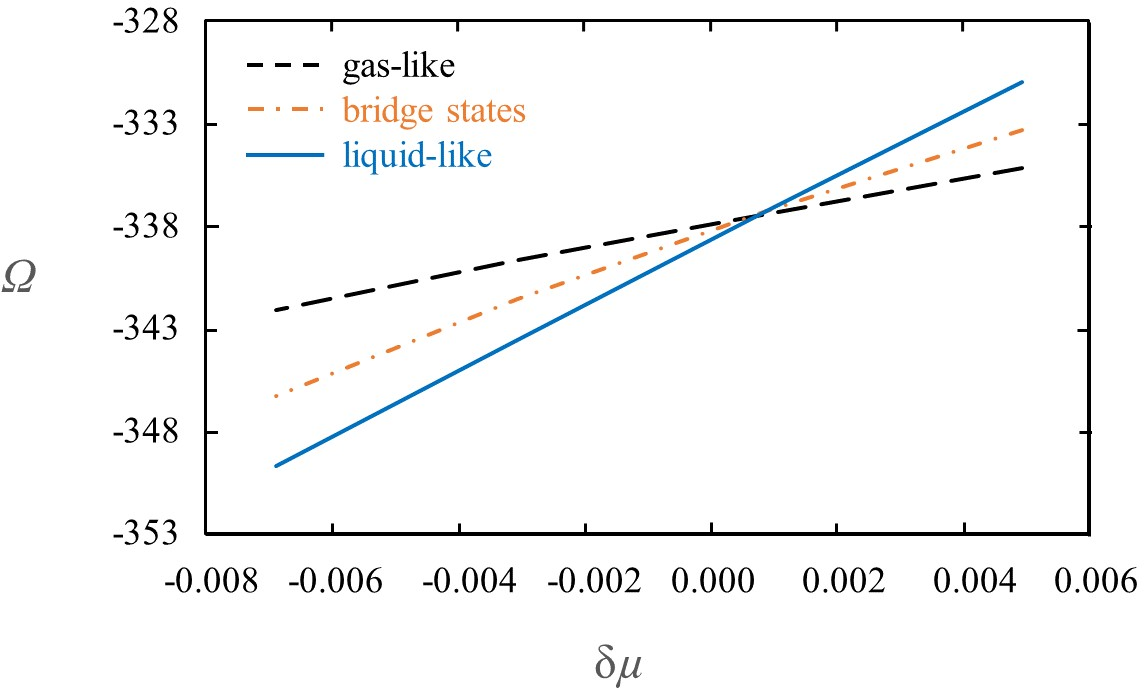}
\caption{DFT results determining the triple point for the maximum contrast slit with $H_1=H_2=40\,\sigma$ and slit width $L=48\,\sigma$. The graph
shows that the grand potential (per unit length, over one cell and in units, $\varepsilon$, of the fluid-fluid potential strength \cite{SP}) as a
function of the undersaturation [$\delta\mu=(\mu_{\rm sat}-\mu)/\varepsilon$] for the gas-like, liquid-like and bridge configurations all intersect
at a value of the chemical potential close to, but slightly away from bulk saturation. The triple point aspect ratio $a_T= 1.2$, below which Cassie's
law breaks down, is slightly above the macroscopic prediction $a_T=1$ in accordance with Eq.~(\ref{aT_meso}).}
\end{figure}

\begin{figure*}
\centerline{\includegraphics[width=5cm]{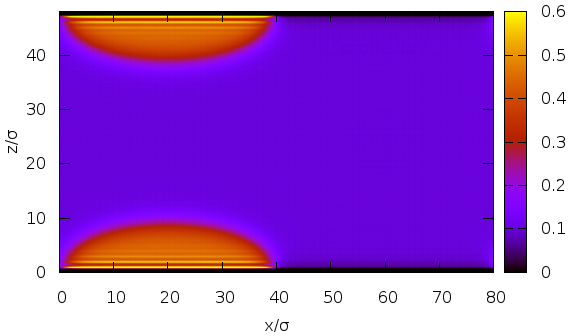} \hspace*{0.5cm} \includegraphics[width=5cm]{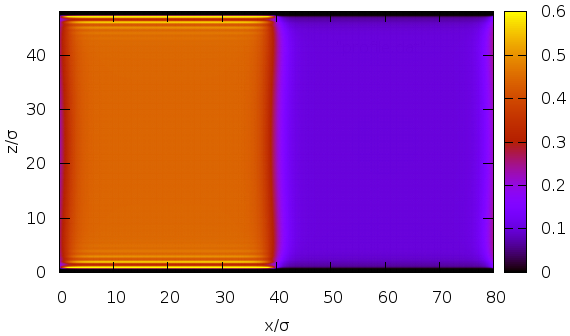} \hspace*{0.5cm} \includegraphics[width=5cm]{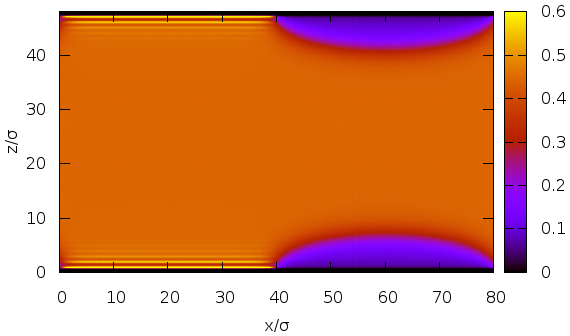}}
\caption{Equilibrium density profiles (over one period) for coexisting states at the triple point for a maximum contrast slit with
$H_1=H_2=40\,\sigma$ and $L=48\,\sigma$. Left and right correspond to gas-like and liquid-like phases for which liquid droplet and gas bubbles are
visible, while the centre plot corresponds to the bridge state with a near flat meniscus.}
\end{figure*}

\begin{figure}
\includegraphics[width=8cm]{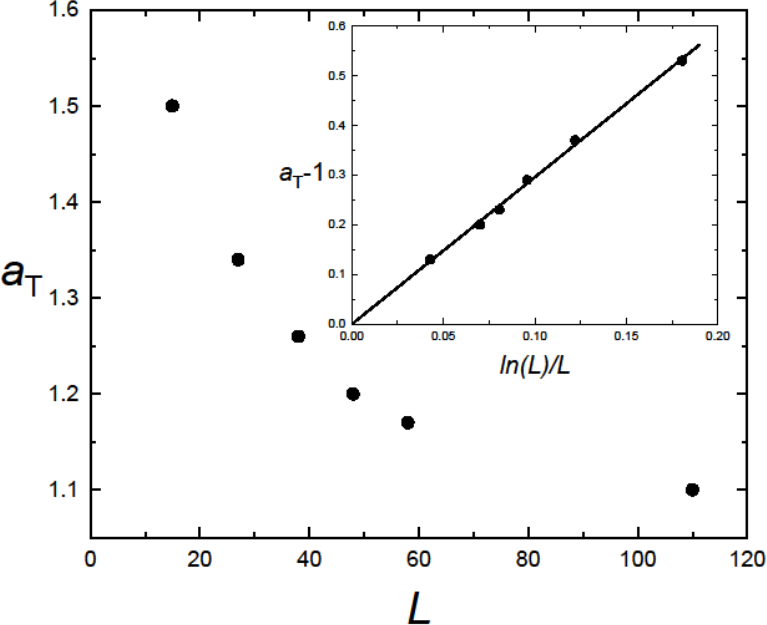}
\caption{DFT results of the triple point aspect ratio of a maximum contrast slit for increasing slit width (in units of $\sigma$) showing the
approach to the macroscopic result $a_T=1$. The inset shows that the mesoscopic result (\ref{aT_meso}) is accurate down to molecularly narrow slits
with $L\approx 15\,\sigma$.} \label{fig4}
\end{figure}

These macroscopic arguments do not allow for the direct influence of dispersion forces. Recall that for a homogeneous slit, they may be safely
ignored for partial wetting but lead to the Derjaguin correction (\ref{Derjaguin}) for complete wetting. The situation is somewhat richer for the
patterned slit. If both materials are partially wet, then, as for the homogeneous case, we anticipate that the Kelvin equations for one and two step
condensation remain accurate down to molecularly narrow slits. However, if one of the materials is wet (or dry) then the influence of the long-range
forces become important at the mesoscopic scale. This is most transparent when we consider the maximum contrast slit described above where material 1
is completely wet and material 2 is completely dry.  In the gas-like phase each stripe of material 1 is wet by a drop of liquid and, in the
liquid-like phase, each stripe of material 2 is wet by a bubble of gas. The volumes of these, which would become macroscopic as $H_1$, $H_2$ and $L$
increase need to be taken into account when we consider, for example, the value of the pressure shift $\delta p_T$ at the triple point. The shape of
these drops and bubbles is determined by the intermolecular forces and can be calculated using  interfacial Hamiltonian methods \cite{posp_stripe}.
The pressure shift at the triple point for the maximum contrast slit is then modified from (\ref{kelvin_cassie})  to
\begin{equation}
\delta p_T=\frac{2\gamma\cos\theta_{\rm cas}}{L-\frac{\pi}{4}(f_1 l_D+f_2l_B)}\,, \label{Derjaguin2}
\end{equation}
where $l_D= (A_1/2\gamma)^{1/4}\sqrt{H_1}$ is the maximum thickness of the wetting drop and $l_B= (A_2/2\gamma)^{1/4}\sqrt{H_2}$ is the maximum
thickness of the drying bubble. Here $A_1$ and $A_2$ are the Hamaker constants for each material. Since at the triple point $H_1$ and $H_2$ are both
of order $L$, the reduction in the effective slit width is \emph{greater} than that occurring  in a homogenous system (since at condensation
$\ell\sim L^{1/3}$). Note that as $H_2$ is reduced to a value $\propto \ln H_1$ the drops coalesce to cover the whole surface corresponding to a
first-order wetting transition at which $\theta_{\rm cas}$ vanishes \cite{posp_bridge}. For smaller values of $H_2$ both walls are completely wet and
the location of the single step condensation is described by the usual Derjaguin correction (\ref{Derjaguin}).

The location of the triple point for the maximum contrast slit is most subtle when the area fractions are equal, $f_1=f_2=1/2$, since in that case
the wall is overall neutral, $\theta_{\rm cas}=\pi/2$. The Kelvin-Cassie equation (\ref{kelvin_cassie}) predicts that single step condensation occurs
at $p_{\rm sat}$, while  Eq.~(\ref{aT}) predicts that the triple point occurs for $a_T=1$ (see Fig.~2). This macroscopic prediction for $a_T$ is easy
to understand. The menisci which bound the liquid bridges are flat with a free-energy cost of $2\gamma L$ per unit cell. For the gas-like
(liquid-like) phase this must be compensated by having a drop of liquid (bubble of gas) coat the wet (dry) stripes which carries with it a
macroscopic free-energy cost of $2\gamma H_1$ per unit cell. Thus, purely macroscopically, the triple point must occur for $L=H_1$. When we include
dispersion forces however, the surface free-energy of a liquid drop, and similarly for the gas bubble, contains a Casimir-like contribution
$\sqrt{2A_1/\gamma}\ln\left(H_1/\sigma\right)$ where $\sigma$ is a molecular diameter. Taking these into account determines the
 higher-order contribution to Eq.~(\ref{Derjaguin2}) when the walls are neutral
\begin{equation}
\delta p_{T}=\sqrt{2\gamma}(\sqrt{A_1}-\sqrt{A_2})\frac{\ln L/\sigma}{L^2}\,;\;\;\;\;\; {\rm for}\; \theta_{\rm cas}=\frac{\pi}{2}
\end{equation}
This small shift therefore owes its origin to the difference in the strengths of the dispersion forces; we note that it is larger than the shift
$\delta p_{cc}=2A/L^3$ for a homogeneous slit when $\theta=\pi/2$. Similarly, for the aspect ratio we find
 \begin{equation}
a_T=1+\frac{\sqrt{A_1}+\sqrt{A_2}}{\sqrt{2\gamma}}\frac{\ln L/\sigma}{L}+\cdots\,, \label{aT_meso}
\end{equation}
which approaches unity as the slit width increases.

We have tested the above prediction for the value of $a_T$ for the maximum contrast slit using a microscopic density functional theory (DFT)  which
we use to determine the equilibrium density profiles and free-energies of stable and metastable phases \cite{evans79}. These are obtained by
minimizing a grand potential functional
\begin{equation}
 \Omega[\rho]={\cal F}[\rho]+\int\dd\rr\rhor[V(\rr)-\mu]\,,\label{omega}
\end{equation}
where $V(\rr)$ is the external potential modelling the long-ranged interaction from the patterned walls and ${\cal F}[\rho]$ is the intrinsic
free-energy functional for which we use Rosenfeld's Fundamental Measure Theory \cite{ros} (see Supplemental Material \cite{SP}).
We have determined the phase coexistence for  system sizes ranging from $L=10\,\sigma$ to $L=110\,\sigma$ and for different stripe widths $H_1$
(=$H_2$). Fig.~3 shows the equilibrium grand potential versus the undersaturation close to the triple point for  $H_1=H_2=40\,\sigma$ and
$L=48\,\sigma$. The three coexisting density profiles are shown in Fig.~4. Finally, in Fig.~5 we plot the triple point aspect ratio as a function of
slit width which shows that, as predicted, the value tends to unity as $L$ increases. In the inset we show that the deviation from unity is extremely
well described by the $\ln L/L$ correction in accordance with the prediction (\ref{aT_meso}).

In 3D these results are not affected significantly by thermal fluctuations where their only influence is to round the bridging transitions associated
with two-step condensation over a pressure range $\mathcal{O}({\rm e}^{-\gamma LH/k_BT})$. Thermal fluctuations are much more important in 2D for
systems with short-ranged forces. Here, a simple realization of the neutral wall is an Ising strip of width $L$  where the surface spins are fixed to
be up and down each over a distance $H_1$. Single step condensation between predominately down spin (analogous to gas) and up spin (liquid) phases,
occurs at zero bulk field, $h=0$, rounded over a scale $\mathcal{O}({\rm e}^{-\gamma L/k_B T})$, while bridging occurs away from $h=0$ (except at the
triple point) and is rounded over a scale $\mathcal{O}(1/LH_1)$ \cite{privman}. Interfacial wandering in all three phases leads to similar mesoscopic
corrections to those predicted for dispersion forces. These can be determined using random walk arguments and interfacial models \cite{fisher84,
parry92, jak}. For the gas phase, the entropic repulsion of the interfaces that bound the liquid drops from the walls leads to a partition function
(per unit cell and at $h=0$) given by $Z_g\approx e^{-2\gamma H_1/k_BT}/H_1^3$ and similarly for the liquid phase. For the bridge, on the other hand
$Z_b\approx {\rm e}^{-2\gamma L/k_BT}/L$ arising from the wandering of the two near flat menisci. Balancing the free-energies determines that a
pseudo triple point occurs at $h=0$ when the aspect ratio is $a_T\approx 1+\frac{k_BT}{\gamma}\frac{\ln L}{L}$. This may be checked numerically and
may even be amenable to exact analysis \cite{binder, zuba, abraham07, abraham}.

In this paper, we have shown that the locations of  one-step and two-step capillary condensation in patterned slits are described by Kelvin equations
involving three possible contact angles, only one of which is the  Cassie angle $\theta_{\rm cas}$ for which we give an explicit geometrical
interpretation. However, for narrow slits Cassie's law is broken and two-step condensation is characterized by different angles which arise from
meniscus pinning. The pinning associated with these angles not only determines the phase equilibria but will also have a strong influence on
metastability relevant to experimental studies and may well underpin the phenomena of contact angle hysteresis. Mesoscopic Derjaguin-like corrections
are significantly larger than those for condensation in homogenous slits and are predicted in both 2D and 3D.

 This work was financially supported by the Czech Science Foundation, Project No. 20-14547S, and the European
Union's Horizon 2020 research and innovation program (Project VIMMP: Virtual Materials Marketplace, No. 760907).


\begin{thebibliography}{99}

 \bibitem{hertlein}
  C. Hertlein, L. Helden, A. Gambassi, S, Dietrich, and C. Bechinger, Nature {\bf 451}, 7175 (2008).

\bibitem{fisher}
M. E. Fisher and H. Nakanishi,  J. Chem. Phys. {\bf 75}, 5857 (1981).

\bibitem{evans1}
R. Evans, P. Tarazona and U. Marini Bettolo Marconi, J. Chem. Phys. {\bf 84}, 2376 (1986).

\bibitem{evans2}
R. Evans, J. Phys.: Condens. Matter. {\bf 2}, 8989 (1990).

\bibitem{hend}
D. Henderson, {\it Fundamentals of Inhomogeneous Fluids}, (New York: Dekker,  1992).

\bibitem{parry_cwalls}
A. O. Parry and R. Evans, Phys. Rev. Lett. {64}, 439 (1990).

\bibitem{binder03}
 K. Binder, D. Landau, and M. M\"{u}ller, J. Stat. Phys.  {\bf 110}, 1411 (2003).

\bibitem{bruschi}
 L. Bruschi, G. Mistura, L. Prasetyo, D. D. Do, M. Dipalo, and F. De Angelis, Langmuir {\bf 34}, 106 (2018).

  \bibitem{cassie}
A. B. D. Cassie, Discuss. Faraday Soc. {\bf 3}, 11 (1948).

\bibitem{degennes}
P. de Gennes, F. Brochard-Wyart, and D. Qu\`{e}r\`{e}, {\it Capillarity and Wetting Phenomena: Drops, Bubbles, Pearls, Waves}, Springer New York,
(2013).

 \bibitem{quere}
 D. Qu\`{e}r\`{e}, Annu. Rev. Mater. Res. {\bf 38}, 71 (2008).


\bibitem{thomson}
W. Thomson, Phil. Mag. {\bf 42}, 448 (1871).


\bibitem{derj}
B. V. Derjaguin, Acta Phys. Chem. {\bf 12}, 181 (1940).


 \bibitem{evans85}
R. Evans and U. Marini Bettolo Marconi,  Chem. Phys. Lett.  {\bf 114}, 415 (1985).

\bibitem{lipowsky85}
R. Lipowsky, Phys. Rev. B {\bf 32}, 1731 (1985).

\bibitem{dietrich}
S. Dietrich, in {\it Phase Transitions and Critical Phenomena}, edited by C. Domb and J. L. Lebowitz (Academic, New York, 1988), Vol. 12.

\bibitem{laska}
M. L\'aska, A. O. Parry, and A. Malijevsk\'y, Phys. Rev. Lett. {\bf 124}, 115701 (2020).

\bibitem{chmiel}
G. Chmiel, K. Karykowski A. Patrykiejew, W. R\.{z}ysko and S. Soko\l owski, Mol. Phys. {\bf 81}, 691 (1994).

\bibitem{rocken}
 P. R\"ocken, A. Somoza, P. Tarazona, and G. Findenegg, J. Chem. Phys. {\bf 108},  8689 (1998).

\bibitem{fin_slit}
A. Malijevsk\'y, A. O. Parry, and M. Posp\'\i \v sil, Phys. Rev. E {\bf 96}, 020801(R) (2017).

\bibitem{posp_stripe}
A. Malijevsk\'y, A. O. Parry, and M. Posp\'\i \v sil, Phys. Rev. E {\bf 96}, 032801 (2017).

\bibitem{posp_bridge}
A. Malijevsk\'y, A. O. Parry, and M. Posp\'\i \v sil, Phys. Rev. E {\bf 99}, 042804 (2019).

\bibitem{evans79}
R. Evans, Adv. Phys. {\bf 28}, 143 (1979).

\bibitem{ros}
Y. Rosenfeld,  Phys. Rev. Lett. {\bf 63}, 980 (1989).

\bibitem{SP}
See Supplemental Material for details of the DFT model.

\bibitem{privman}
V. Privman and M. E. Fisher, J. Stat. Phys. {\bf 33}, 385 (1983).

\bibitem{fisher84}
 M. E. Fisher, J. Stat. Phys. {\bf 34}, 667 (1984).

 \bibitem{parry92}
 A. O. Parry and R. Evans, J. Phys. A {\bf 25} 275 (1992).

\bibitem{jak}
P. Jakubczyk, M. Napi\'{o}rkowski, and A. O. Parry, Phys. Rev. E {\bf 74}, 031608 (2006).

\bibitem{binder}
K. Binder, D. Landau, and M. Muller, J. Stat. Phys. {\bf 110}, 1411 (2003).

\bibitem{zuba}
M. Zubaszewska, A. Gendiar and A. Drzewinski, Phys. Rev. E {\bf 86}, 062104 (2012).

\bibitem{abraham07}
D. B. Abraham, F. H. L. Essler, and A. Macio\l ek, Phys. Rev. Lett. {\bf 98}, 170602 (2007).

\bibitem{abraham}
D. B. Abraham and A. Macio\l ek, Phys. Rv. Lett. {\bf 105}, 055701 (2010).






























\end{thebibliography}
\end{document}